\documentclass[amsmath,notitlepage,prx,reprint]{revtex4-2}
\usepackage{braket}
\usepackage{amsmath}
\usepackage{amssymb}
\usepackage{physics}
\usepackage{svg}
\usepackage{graphicx}
\usepackage{hyperref}
\usepackage{bbding}
\hypersetup{
    colorlinks=true,
    linkcolor=blue,
    citecolor=blue,
    urlcolor=blue
}

\newcommand{\e}[1]{\ensuremath{\times{10^{#1}}}}
\newcommand{\ts}[2]{\ensuremath{{#1}_{\text{#2}}}}
\newcommand{\CE}[1]{\ts{#1}{CE}}
\renewcommand{\vec}[1]{\mathbf{#1}}
\newcommand{\figref}[2]{Fig.~\hyperref[#1]{\ref*{#1}(#2)}}
\newcommand{\figureref}[2]{Figure~\hyperref[#1]{\ref*{#1}(#2)}}
	
\begin{document}
	
	\title{Benchmarking exchange-only control of a 48-spin singlet manifold}

	\author{HRL Quantum Team\hyperlink{hrl_authors}{$^\dagger$} and Microsoft Collaborators\hyperlink{msft_authors}{$^\ddag$}}
	\date{\today}
	
	\begin{abstract}
		Exchange-only quantum computing benefits from high fidelity and straightforward control afforded by the exchange interaction.
		However, independently controllable qubits must be encoded into subsystems of at least three electron spins, restricting computation to only a fraction of the available spin Hilbert space.
		The remaining states are treated as leakage and used only transiently in gate sequences.
		By contrast, a quantitative, system-wide measure of exchange-only performance can exploit the full available Hilbert space, including states conventionally treated as leakage.
		Here, we apply this approach to arrays of up to 48 electron spins, accessing a total-spin-zero Hilbert space with dimension exceeding $2^{40}$.
		Measurements of out-of-time-order correlators (OTOCs) reveal rich scrambling dynamics and demonstrate access to regimes relevant to quantum computational advantage.
		Using generalized forms of cross-entropy and mirror randomized benchmarking, we also assess the aggregate performance of the full exchange-only system per fundamental two-body interaction: the two-spin exchange.
		We obtain an effective system-level error of $3\e{-4}$ per exchange, incorporating the complete experimental control sequence and all associated noise sources.
		This value is up to an order of magnitude lower than those reported from comparable benchmarks on other platforms at the time of writing.
	\end{abstract}
	
	\maketitle
	
	\section{Introduction}
	
	Exchange-only (EO) quantum computers perform quantum computation with only voltage-controlled spin-spin exchange interactions.  
	One motivation is to simplify controller hardware, enabling cryogenic control, as we demonstrated in Ref.~\cite{tn3_paper} and extend here.
	A second reason is that the fidelity of the exchange interaction can be made extremely high, as evidenced by studies of miscalibration-compensating dynamical decoupling, which estimate charge-noise-limited errors as low as $10^{-5}$ per exchange operation~\cite{Bo2024}.   
	However, with only exchange and uniform magnetic fields, single-spin Pauli gates are disallowed by symmetry.  
	Universal quantum computing is made possible by grouping three spins into a ``decoherence-free subsystem'' (DFS) qubit~\cite{zanardi97,bacon_universal_2000,kempe_theory_2001,divincenzo_universal_2000,Fong2011,Bo2024,weinstein_universal_2023,tn3_paper}, reducing the total state dimension of $N$ spins from $2^N$ states to the exponentially smaller dimension of $2^{N/3}$. 
	
\begin{figure*}[t]
		\centering
		\includegraphics[width=\textwidth]{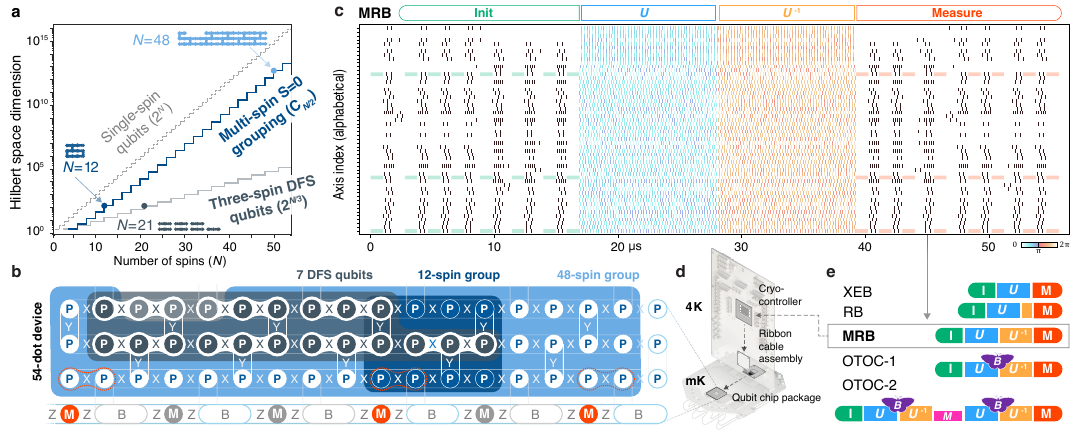}
		\caption{
			\label{fig:intro}
			\textbf{a} Scaling of the quantum Hilbert space size with the number of spins for single-spin qubits, three-spin DFS qubits, and the $S=0$ operation mode studied in this work.
			\textbf{b} Flexibility of exchange-only operation on the qubit chip, with device subgraphs representing 7-qubit DFS, 12-spin $S=0$, and 48-spin $S=0$ highlighted in various shades of blue. The three measurement sites used in this work are highlighted in red.
			Particular unused spins and measurement dots were judged more challenging to calibrate due to disorder present in this system.  
			\textbf{c} Example of an exchange program for mirror randomized benchmarking (MRB) executed in the $S=0$ operation mode.
			Exchange angle magnitudes are given in color, with $\pi$ pulses (spin swap) in black, angles less than $\pi$ in shades of blue, and angles greater than $\pi$ in shades of red.  
			The top shows a schematic of the program in terms of four sections: spin-pair initialization (Init), a random sequence ($U$), an inversion of that random sequence ($U^{-1}$), and spin-pair measurement (Measure).
			\textbf{d} Schematic of the QPU, including the cryo-controller, which plays the exchange sequence shown in (c), the 4K-to-mK ribbon cable, and the qubit package, which holds the 54-spin qubit chip shown in (b).
			\textbf{e} Schematics of benchmarking and out-of-time-order correlator (OTOC) sequences: cross-entropy benchmarking (XEB), randomized benchmarking (RB), MRB, first-order OTOC (OTOC-1), and second-order OTOC (OTOC-2).
		}
	\end{figure*}

	The 3-spin DFS, however, is but one of many encodings available for exchange-only control.  
	Conventionally, a DFS may be defined by $N$ spins in the total spin angular momentum $S=0$ state for even $N$ or an $S=1/2$ state for odd $N$.
	The dimension $d$ of the even-$N$ $S=0$ subspace is the Catalan number of half the spin count, $d = C_{N/2}$, and the size of the odd-$N$ $S=1/2$ subsystem is $d=C_{(N+1)/2}$.
	For reasonably large $N$, $C_{N/2}$ is well-approximated by $2^{N}/(N^{3/2}\sqrt{\pi})$, meaning the controllable space of $N$ spins is equivalent in size to about $N-(3/2)\log_{2} N$ effective qubits.
	This scaling is shown in \figref{fig:intro}{a}.  
	For the largest grouping explored here, $N=48$ spins in \figref{fig:intro}{b}, 
	$d$ is slightly larger than that provided by 40 qubits ($C_{24} \approx 2^{40}$).  
	As observed long ago~\cite{kempe_theory_2001}, exchange-only quantum computing inside a single $N$-spin DFS can support a Hilbert space that scales with $N$ at almost the same rate as a quantum computer made of single-spin qubits!
	
	When we operate the system with such large $S=0$ encodings, though, the cost is an absence of well-defined local qubits, leading to unresolved questions about how to synthesize quantum programs or establish fault-tolerant error correction protocols.
	Universal control over the available space, using $S$-conserving exchange operations of variable strength between spins, is certainly available in principle.  
	While the design of \emph{arbitrary} unitary programs with finite sequences of exchange operations in these subspaces is less developed than circuit synthesis for standard qubits, we can at least develop Haar-random distributions of unitaries, or unitaries that simulate physically interesting Hamiltonians related to Heisenberg exchange~\cite{Hoffman2026}.  
	The control landscape is therefore useful, even if it is not that of conventional, local-qubit-based quantum computing. 
	Accordingly, throughout this paper the constituent two-level systems we control are called ``spins,'' not ``qubits.''  
	In exchange-only operation, these spins are initialized in entangled pairs (as spin singlets $(\ket{\uparrow\downarrow}-\ket{\downarrow\uparrow})/\sqrt{2}$, i.e., $\Psi^-$ Bell states), and every exchange operation, other than full-spin swaps, potentially grows that initial entanglement.  
	In this sense, there is no notion of a single-spin gate; \emph{all} gates are two-spin, and \emph{all} available states have some level of spin entanglement.

	Most recent quantum advantage and benchmarking experiments define their circuits in terms of the familiar universal gate sets of qubits: tensor products of Pauli operators and the Clifford operations that normalize their group (CZ, H, and so on).   
	With most such operations lacking in $S=0$ subspaces, one might worry that highly specialized compilation methods are needed.
	However, at least two system-wide experiments on quantum computers, random-circuit sampling (RCS) and out-of-time-order correlators (OTOCs), are easy to define in $S=0$ operation.
	The case of RCS follows precedent from Google Quantum AI, which relied on RCS in an early device lacking CZ entanglers to demonstrate ``quantum supremacy''~\cite{Arute2019}.  
	In both that result and ours, the entanglers are well understood and well calibrated; they just do not adhere to the Clifford group structure relied upon by many quantum computing benchmarks.
	In our case, the non-Clifford operations are calibrated exchange interactions, i.e., well-characterized fractional spin-swaps, with known but randomly selected magnitudes [\figref{fig:intro}{c}].  
	
	The benchmarking and OTOC experiments studied here [\figref{fig:intro}{e}] were performed on a single quantum processing unit (QPU) of the type described in Ref.~\onlinecite{tn3_paper}.
	The qubit chip [\figref{fig:intro}{b}] contains 54 quantum dots, 63 controllable interdot exchange axes, and six state preparation-and-measurement (SPAM) sites along the bottom rail.
	Of these components, we use up to 48 dots, 54 axes, and three SPAM sites for demonstrations here.
	Voltages applied to the plunger electrodes are tuned to confine a single electron in each active quantum dot, while voltages applied to the exchange electrodes control the exchange coupling between neighboring dots.
	The exchange interactions are activated by time-calibrated DC voltage pulses generated by a cryogenic CMOS controller operating inside the dilution refrigerator at the 4-K stage, shown in \figref{fig:intro}{d}.

	In these experiments, the two-spin exchange interaction serves as the native two-body operation.
	Each exchange pulse implements a calibrated fractional spin swap, and sequences of these operations spread entanglement and
	scramble quantum information within the accessible Hilbert space.
	In this sense, exchange operations are analogous to two-qubit operations in conventional many-qubit architectures.
	We demonstrate excellent performance of the two-spin exchange interaction (including magnetic and system miscalibration effects, unlike in Ref.~\onlinecite{Bo2024}), and execute accurate, high-depth calculations of OTOCs.
	The loss in fidelity per exchange, $\ts\epsilon{F} = 3.2\e{-4}$, compares favorably to errors $>10^{-3}$ per two-qubit gate currently reported for other quantum computing platforms~\cite{Chen2024, Madzik2025, Abanin2025, Cosentino2026, Dasu2026, AtomComputing2026, Evered2026}.
	
	In Sec.~\ref{sec:s0}, we introduce practical details of operating in the total-spin-zero space. 
	Sec.~\ref{sec:benchmarks} presents results of XEB and MRB benchmarking. 
	Sec.~\ref{sec:otoc-1} and Sec.~\ref{sec:otoc-2} contain various OTOC experiments. 
	Finally, in Sec.~\ref{sec:discussion}, we summarize our results in context with recent experiments in the literature.

	\section{$S=0$ operation}
	\label{sec:s0}

	Operation in the $S=0$ subspace begins from a known total-spin-zero state, for which we use the simplest choice: a product of contiguous nearest-neighbor singlets covering the active array.
	We prepare singlets at the SPAM sites, marked ``M'' in \figref{fig:intro}{b}, and route them into the array using $\pi$-exchange pulses.
	The routing depends on the size and connectivity of the active array.
	For measurement, the routing is reversed and each pair is measured in the singlet--triplet basis, with outcomes recorded as ``0'' and ``1'', respectively.
	Further details are provided in Appendix~\ref{appendix:spam}.

	Starting from the initialized singlet-product state, we construct native exchange sequences for either random many-body evolution or Trotterized evolution under a target Hamiltonian.
	Specifically, ``random'' refers to sequences drawn from the native ensemble described below, whose statistics approach those of Haar-random unitaries~\cite{Oliveira2007, Weinstein2008, Brown2013} within the accessible $S=0$ subspace as the sequence depth increases.
	We use random sequences for cross-entropy benchmarking, mirror randomized benchmarking, and the random OTOC protocols, whereas Trotterized sequences are used to emulate a Heisenberg chain and probe its dynamics through OTOCs.
	In the example sequence diagram of \figref{fig:intro}{c}, and throughout this paper, we denote the forward evolution by $U$ and its inverse by $U^{-1}$. 

	In the case of random evolution, exchange operations are arranged in a brickwork pattern, with each angle sampled from the discrete set $\left\{\pi/5, 2\pi/5, 3\pi/5, 4\pi/5, \sqrt{2}\right\}$.
	We implement $U^{-1}$ by reversing the sequence and replacing each exchange angle $\theta$ with its $2\pi$ complement, $2\pi-\theta$.
	All random angles are sampled \emph{a priori} rather than generated on-chip.
	Resolving an error per exchange at the $10^{-4}$ scale requires sequences containing more than $10{,}000$ exchange operations.
	Because such sequences exceed the finite instruction memory of the cryogenic controller, we employ a recursive scheduling strategy in which long random sequences are assembled hierarchically from shorter, precompiled random subroutines (see Appendix~\ref{appendix:rcs}).
	This recursive construction enables substantially deeper sequences within the available controller memory, but repeated calls to precompiled subroutines introduce temporal structure relative to fully independently sampled brickwork sequences.
	Nevertheless, we find that the subroutines retain substantial internal randomness, and the resulting sequences exhibit comparable randomness at matched exchange count.

	The efficiency of the random sequence, defined as the depth-dependent convergence toward Haar-random statistics, depends on the connectivity of the device and the recipe for the brickwork pattern.
	On strictly one-dimensional spin arrays, exchange brickwork sequences restricted to the $S=0$ sector are less efficient per exchange than analogous brickwork circuits on one-dimensional qubit arrays, although the relative difference diminishes with increasing system size, consistent with the Hilbert-space comparison in~\figref{fig:intro}{a}.
	But such architecture dependence is not unique to exchange-only dynamics.
	For instance, all-to-all connectivity leads to more efficient brickwork circuits than geometrically local connectivity~\cite{Haferkamp2021, Schuster2025, decross_computational_2025}.
	We therefore focus not on the efficiency of our random sequences, but on the aggregate system performance expressed as an effective error per fundamental two-body interaction.
	
	\section{Benchmarks}
	\label{sec:benchmarks}
	
	\begin{figure}[t]
		\centering
		\includegraphics[width=\columnwidth]{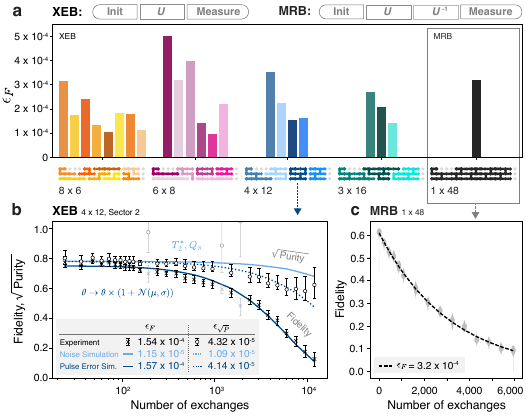}
		\caption{
			\label{fig:benchmarks}
			\textbf{a} Loss in fidelity per exchange, $\epsilon_F$, for 48-spin benchmarks in the $S=0$ operation mode.
			Parallel XEB experiments, of varying spin counts (color) and subgraphs (shade), are shown alongside a 48-spin MRB experiment, with the spin subgraph used for each data point drawn underneath with a matching color.
			Uncertainties in the fitted $\epsilon_F$ are all less than $10^{-5}$ and have been omitted.
			\textbf{b} XEB fidelity versus exchange pulse count for a 12-spin XEB experiment, given by the black crosses.
			The open circles represent the square root of purity, which measures the loss in coherence, such as from averaging over stochastic control errors or environmental decoherence.
			A simulation that includes the microscopic noise terms $Q_{\pi}$ and $T_2^*$, as measured on individual axes and spins of the same device, is shown in light blue.
			A second simulation, which uses a heuristic pulse-error model fitted to the experimental data, is shown in dark blue.
			The losses in XEB fidelity per exchange ($\epsilon_F$) and in square-root purity per exchange ($\epsilon_{\sqrt{\mathcal{P}}}$) for the experiment and both simulations are given in the inset table.
			The experimental error bars are obtained from bootstrapping over subsamples of shots and sequences at each exchange count.
			\textbf{c} Violin plot of MRB fidelity versus pulse count, with an exponential fit (dashed black line) giving the loss in fidelity per exchange pulse ($\epsilon_F$) of $3.20 \times 10^{-4}$. The fit uncertainty is $5.5 \times 10^{-6}$.
		}
	\end{figure}

	\subsection{Cross-Entropy Benchmarking (XEB)}

	We first characterize random exchange evolution using a generalized form of cross-entropy benchmarking (XEB).
	For each random exchange sequence, we compare the measured singlet--triplet output probabilities with the ideal probabilities obtained from a noiseless classical simulation of the same sequence.
	Normalizing by the ideal probability variance, the covariance between the measured and ideal probabilities estimates the sequence fidelity, while the variance of the measured probabilities over the random sequence ensemble estimates the output-state purity.

	Applying XEB directly to the full 48-spin array is limited by the coarse-grained nature of Pauli spin-blockade readout.
	Each spin pair has one singlet and three triplet states, but the measurement resolves only a binary singlet--triplet outcome.
	A bitstring containing $N_T$ triplet outcomes therefore corresponds to a measurement subspace of dimension $3^{N_T}$ in the full spin Hilbert space.
	Within the $S=0$ sector, the relevant rank is the Riordan number $R_{N_T}$, which counts the ways that $N_T$ spin-1 triplets can combine to total spin zero.
	As this rank increases, the random-sequence fluctuations of the ideal outcome probabilities are suppressed, reducing the statistical sensitivity of XEB.
	For example, a measurement in which the first 20 of the 24 measured spin pairs yield triplet outcomes and the last 4 yield singlets combines $3^{20} \cdot 1^4 \approx3.5\times10^{9}$ different states into a single measured bitstring.
	Even when restricted to $S=0$, this outcome has rank $R_{20}=13{,}393{,}689$, making direct 48-spin XEB statistically impractical.

	To retain resolvable output-probability fluctuations while benchmarking all 48 spins simultaneously, we partition the array into disjoint subgraphs of $N\leq16$ spins and perform XEB independently on each subgraph in parallel [\figref{fig:benchmarks}{a}].
	We label each partition by $n_{\mathrm{g}}\times N$, where $n_{\mathrm{g}}$ is the number of subgraphs and $N$ is the number of spins in each subgraph.
	For example, $8\times6$ denotes eight six-spin subgraphs, whereas $3\times16$ denotes three sixteen-spin subgraphs.
	We use four such partitions, each spanning all 48 spins.
	For each subgraph, we fit the fidelity and purity decays as an exponential function of exchange count.
	Averaging the fitted XEB fidelity loss (error) per exchange operation across all subgraphs yields a device-wide average exchange error of $\epsilon_F = 2.2 \times 10^{-4}$.
	Note that, for XEB and MRB, $\epsilon_F$ is analogous to an error per layered gate (EPLG) reported by layer fidelity benchmarks~\cite{McKay2023}.

	The same XEB data also provides a measure of output-state purity through speckle purity benchmarking (SPB) \cite{van_enk_measuring_2012, Arute2019}.
	SPB compares the variance of the measured output probabilities across random sequences with the corresponding ideal variance.
	Deterministic coherent errors can reduce fidelity while preserving the purity of each output state, whereas averaging over stochastic control errors or environmental decoherence generally produces a mixed state and reduces both fidelity and purity.
	Their relative decay therefore provides information about the character of the underlying errors, although it does not uniquely identify individual noise mechanisms.
	For $S=0$ experiments, the ideal probability variance depends on measurement degeneracy.
	We account for this by grouping outcomes with the same number of measured triplets, $N_T$, and normalizing their variance by the corresponding value that would be achieved under noiseless operation.
	We fit the resulting square-root purity, $\sqrt{\mathcal{P}}$, and report its loss per exchange, $\epsilon_{\sqrt{\mathcal{P}}}$.
	This estimator has limitations, as it is highly sensitive to correlated measurement errors, and only some of the XEB subgraphs had the measurement stability required for extraction of $\epsilon_{\sqrt{\mathcal{P}}}$ (see Appendix~\ref{appendix:xeb}).

	In \figref{fig:benchmarks}{b}, we examine the dependence of the XEB fidelity and square-root purity on exchange count for a twelve-spin subgraph.
	The fidelity decays substantially faster than the square-root purity, producing a clear separation between the extracted errors per exchange: $\epsilon_F$ and $\epsilon_{\sqrt{\mathcal{P}}}$.
	This indicates that coherent (shot-independent) error is a large contributor to this sector's XEB infidelity.
	A noise-aware simulation based on independently measured component-level parameters, namely the exchange quality factor $Q_{\pi}$ and idle dephasing time $T_2^*$ reported in Appendix~\ref{appendix:characterization}, predicts an $\epsilon_F$ more than an order of magnitude smaller than the measured value.
	The simulation also predicts $\epsilon_F \approx \epsilon_{\sqrt{\mathcal{P}}}$, indicating little coherent contribution from the modeled error sources.
	The magnitude and character of the observed error suggest that device performance is limited by coherent and incoherent pulse errors that do not occur during single-component characterization.
	We suspect nonidealities in the cryo-controller and signal lines are dominant contributors.
		
	To better understand the device performance, we introduce a simple phenomenological model of exchange-angle errors.
	As pulse distortions can modify integrated pulse angles, the model transforms angles according to
	\begin{equation}
		\CE{f}(\theta) = \theta \times [ 1 + \mathcal{N}(\CE\mu, \CE\sigma)],
		\label{eq:pulse_errors}
	\end{equation}
	where $\mathcal{N}$ is a normally distributed random variable with mean $\CE\mu$, a systematic multiplicative offset, and standard deviation $\CE\sigma$, the magnitude of stochastic angle fluctuations.
	We emphasize that Eq.~\ref{eq:pulse_errors} is not intended to serve as a realistic description of the pulse errors on the device.
	Rather, it provides an estimate of the effective pulse-error magnitudes required to reproduce the XEB data.
	A simulation that uses Eq.~\ref{eq:pulse_errors} with a coherent offset of $\CE\mu = 7.2 \times 10^{-3}$ and incoherent noise of $\CE\sigma = 8.4 \times 10^{-3}$ produces the blue traces in \figref{fig:benchmarks}{b}, matching the experimental rates of $\ts\epsilon{F}$ and $\epsilon_{\sqrt{\mathcal{P}}}$ for this twelve-spin subgraph.
	
	\subsection{Mirror Randomized Benchmarking (MRB)}

	We next benchmark the full 48-spin array using exchange-only mirror randomized benchmarking (MRB).
	Each experiment begins in the all-singlet state, applies a random exchange evolution $U$, and then applies its inverse $U^{-1}$.
	In the absence of error, the sequence returns deterministically to the initial state.
	We therefore define the benchmark as the probability of observing all 24 spin pairs as singlets.
	Because the all-singlet outcome is nondegenerate, we avoid the measurement degeneracy that limits XEB.
	Conventional MRB combines outcomes at different Hamming distances from the initial state into an effective polarization~\cite{Proctor2022MRB}.
	Here, because each triplet outcome corresponds to a degenerate measurement subspace, we instead use the exact-return probability.
	This choice avoids assigning weights to unresolved triplet outcomes and provides a more stringent metric, as every outcome other than the all-singlet state is counted as a failure.
	
	\figref{fig:benchmarks}{c} shows the all-singlet return probability as a function of exchange count.
	Fitting its decay yields an effective system-level error of
	$\epsilon_F=3.2\times10^{-4}$ per exchange.
	Because the applied angles (exchange pulse durations) in $U^{-1}$ are larger than those in $U$ [\figref{fig:intro}{c}], this result is consistent with the fidelity-derived error obtained from XEB (which only utilizes $U$).

	\section{OTOC-1}
	\label{sec:otoc-1}

	Having characterized system-wide performance through benchmarking, we next use the first-order out-of-time-order correlator (OTOC-1) to probe information spreading within the $S=0$ subspace under exchange dynamics~\cite{Markus2018,Dag2019,Wei2019,Abanin2025}.
	In this section, we study two protocols: random OTOC-1, for which $U$ is generated by native random exchange sequences, and spin-chain OTOC-1, for which $U$ approximates time evolution under a Heisenberg spin-chain Hamiltonian through Trotterization.

	The primary difference between random OTOC-1 experiments and MRB is that OTOC-1 introduces a ``butterfly'' perturbation between the forward evolution $U$ and its inverse.
	In particular, OTOC-1 probes how evolution under $U$ affects the sensitivity of a chosen observable to an unmirrored exchange operation on a ``butterfly pair.''
	Here, we use the singlet--triplet observable of a designated ``measurement pair,'' whose expectation value is determined by the probabilities of singlet and triplet outcomes.
	Let $M$ denote a $\pi$-exchange operator on the measurement pair and $B$ a $\theta_B$-exchange operator on the butterfly pair.
	Importantly, $M$ is both Hermitian and unitary, $M^\dagger=M^{-1}=M$, and hence serves both as a self-inverse exchange operation within an OTOC sequence and as an observable whose expectation value is obtained from singlet--triplet readout ($\pm 1$ eigenvalues).
	These properties are also used in the random OTOC-2 protocol, where $M$ is both applied within the echo sequence and measured at the sequence's end.

	The OTOC-1 sequence evolves the system forward under $U$, applies $B$, and then reverses the evolution with $U^{-1}$.
	The sequence $U$ can represent time evolution under an unspecified Hamiltonian, with the sequence implementing the Heisenberg-picture butterfly operator $B(t)=U^{-1}(t)BU(t)$.
	OTOC-1 is then
	\begin{equation}
		\mathcal{C}_{\psi_0}^{(2)}(t)
		=
		\bra{\psi_0}
		B^\dagger(t)M B(t)M
		\ket{\psi_0}
		=
		2p_S(t)-1,
		\label{eq:otoc1}
	\end{equation}
	where $\ket{\psi_0}$ is the initialized singlet-product state and $p_S(t)$ is the final singlet-return probability of the measured pair.
	The second equality follows from $M = I - 2P_s$, where $P_s$ is the singlet projector of the measured pair, together with $M\ket{\psi_0} = -\ket{\psi_0}$.
	Hence, the rightmost $M$ does not need to be applied physically in the OTOC-1 sequence.
	Equivalently, in terms of commutators,
	\begin{equation}
	    1-p_S(t)
	    =
	    \frac{1}{4}
	    \bra{\psi_0}
	    [B(t),M]^\dagger[B(t),M]
		\ket{\psi_0}.
	    \label{eq:otoc1_commutator}
	\end{equation}
	We therefore present the OTOC-1 results directly in terms of $p_S$.
	In the absence of noise, a decrease in $p_S$ indicates that information has spread from the butterfly pair toward the measured pair, causing $B(t)$ and $M$ to no longer commute.

	\subsection{Random OTOC-1}

\begin{figure}
		\centering
		\includegraphics[width=\columnwidth]{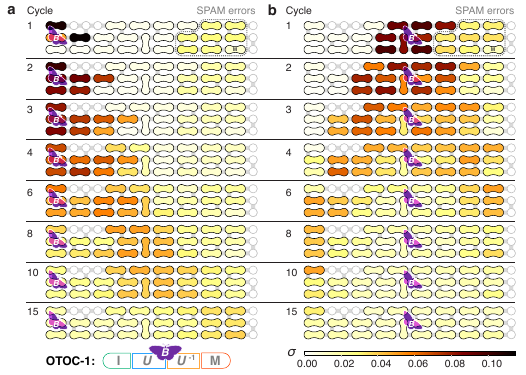}
		\caption{
			\label{fig:otoc1}
			\textbf{a} Standard deviation in singlet-return probabilities ($\sigma$) across 100 different OTOC-1 sequences on 48 spins, with a $\pi/2$ butterfly exchange applied on the left side of the device, as a function of cycle count (1 to 15).
			\textbf{b} Same as (a) but with a butterfly exchange in the middle of the device.
		}
	\end{figure}

	For random OTOC-1, we sample 100 realizations of $U(t)$ at each sequence depth and consider two choices of butterfly pair.
	Depth is measured in cycles, where each cycle activates every available exchange once.
	For the 48-spin array, this corresponds to 54 exchange interactions per cycle.
	Fig.~\ref{fig:otoc1} shows the standard deviation $\sigma$ of the singlet-return probability $p_S$ across the 100 random sequences as a function of cycles in $U$.

	In Fig.~\ref{fig:otoc1}, a large $\sigma$ indicates that the singlet-return probability of the measured pair depends strongly on the particular realization of $U$.
	If the exchange dynamics under $U$ do not propagate the effect of the butterfly operation to the measured pair, the ideal inverse evolution $U^{-1}$ refocuses that pair to its initial singlet state.
	Therefore, the cycle dependence of $\sigma$ can be interpreted as a ``wavefront,'' in which the spin pairs that have the largest $\sigma$ sit on the cusp of a wave of information spreading from the butterfly location through the device due to entanglement propagated by the exchange interactions in $U$~\cite{Abanin2025}.
	Note the slightly elevated $\sigma$ for eight spin pairs on the right side of the device, all of which are measured out of the rightmost measurement site.
	This measurement site experienced periods of large shifts in its Pauli spin-blockade readout current, which we attribute to a local charge instability.
	
	\subsection{Spin-chain OTOC-1}
	\label{sec:spin_chain}

\begin{figure*}
		\centering
		\includegraphics[width=\textwidth]{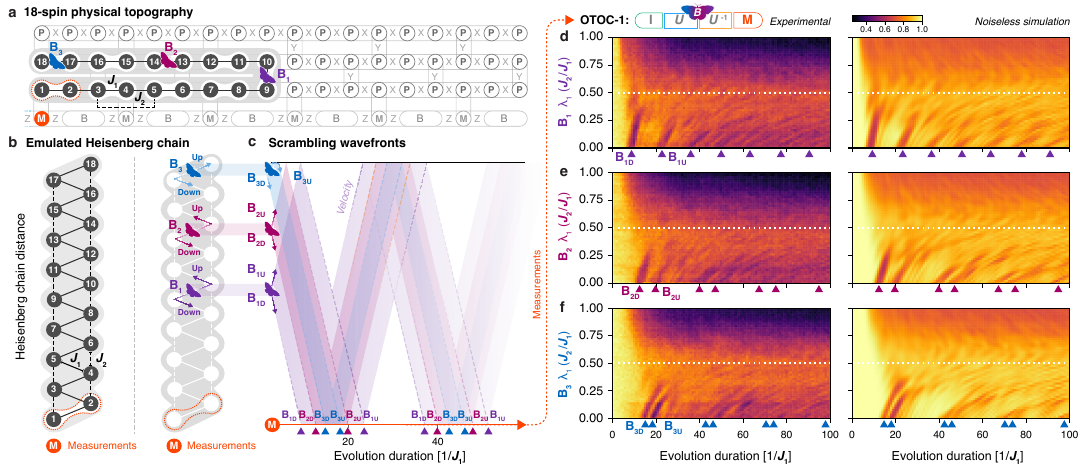}
		\caption{
			\label{fig:heisenberg}
			\textbf{a} Physical locations of the 18 spins used in the OTOC-1 calculation of a Trotterized $J_1$-$J_2$ isotropic 1D Heisenberg spin-chain Hamiltonian.
			Three different butterfly locations, $B_1$, $B_2$, and $B_3$, are indicated at various distances from the measurement spin pair, which is outlined in red.
			\textbf{b} Virtual locations of the 18 spins, as well as a depiction of the two spin-coupling terms $J_1$ and $J_2$. 
			Neighboring sites are staggered such that the $J_1$ interaction forms a diagonal truss between neighboring spins, while the $J_2$ interaction forms a vertical connection between next-nearest neighbors.
			The three butterfly locations now indicate their position on the virtual spin chain.
			\textbf{c} Diagram of the butterfly wavefront dynamics on the 1D spin chain.
			The initially downward-traveling wave from each butterfly ($B_{iD}$) eventually reaches the measurement location, while the upward-traveling wave ($B_{iU}$) must first reflect off the far boundary before reaching the measurement site.
			\textbf{d--f} Experimental measurements (left) and noiseless simulations (right) of the OTOC-1 sequence for three butterfly locations, with the measurement outcome shown in color as a function of the duration of $U$ (in units of $1/J_1$) and the ratio of $J_2$ to $J_1$ ($\lambda_1$).
			Indicated at the bottom of each panel are the expected times at which the waves reach the measurement site, with their spacing matching the horizontal axis of (c).
		}
	\end{figure*}

	We next apply OTOC-1 to Trotterized evolution under a target spin Hamiltonian rather than to random dynamics.
	The $SU(2)$-invariant nature of $S=0$ operation allows for the implementation of many physically motivated models which possess that symmetry.
	Here we consider the isotropic Heisenberg model of magnetic frustration on a spin chain,
	\begin{equation}
		\label{eq:heisenberg}
		H = J_1 \sum_{\langle i, j \rangle} \vec{S}_i \cdot \vec{S}_j + J_2 \sum_{\langle\!\langle i, j \rangle\!\rangle} \vec{S}_i \cdot \vec{S}_j,
	\end{equation}
	where $\langle i,j \rangle$ denotes nearest neighbors on the spin chain and $\langle\!\langle i,j \rangle\!\rangle$ denotes next-nearest neighbors [\figref{fig:heisenberg}{b}].  The parameters $J_1$ and $J_2$ are positive, indicating antiferromagnetism.
	We parameterize the relative strength of the frustrating next-nearest-neighbor interaction by the dimensionless ratio $\lambda_1=J_2/J_1$.
	Note that singlet--triplet readout directly probes the local spin correlation, $p_S=1/4-\langle \vec{S}_i \cdot \vec{S}_j \rangle$, a key observable in frustrated magnetic systems.
	
	An exact simulation of the Hamiltonian in Eq.~\ref{eq:heisenberg} would require continuous exchange interactions between all nearest and next-nearest neighbors, in a manner similar to an analog quantum computer~\cite{Deshpande2026,Andersen2026}.
	As such, it cannot be directly implemented on our device, where even nearest-neighbor simultaneous exchange interactions are difficult to calibrate~\cite{Broz2026}.
	Instead, one must approximate time evolution under Eq.~\ref{eq:heisenberg} by grouping terms into sets that can occur in parallel and applying a Trotter approximation which evolves the different subsets of the Hamiltonian for a duration $\Delta t$.
	For the 18-spin chain in \figref{fig:heisenberg}{a}, an exchange schedule inspired by the NZ1 dynamical-decoupling sequence implements all necessary $J_1$ and $J_2$ interactions~\cite{Bo2024}.
	NZ1 completely permutes spins within a three-spin grouping, allowing all nearest- and next-nearest-neighbor interactions to occur while minimizing accumulated phase errors due to local magnetic gradients.
	We have not attempted to optimize the overall fidelity of a quantum simulation of Eq.~\ref{eq:heisenberg}, and a study of the amount of Trotter error introduced by our choice of $\Delta t J_1 = \pi/3$ is provided in Appendix~\ref{appendix:trotterization}.
	
	In \figref{fig:heisenberg}{d--f}, we report the singlet-return probability on a spin pair at one end of the spin chain after application of an OTOC-1 sequence for three different butterfly locations.
	Sweeping the magnitude of $J_2$ and the total $U$ evolution duration reveals a rich structure of butterfly-induced wavefronts.
	To the right of each experimental result is the outcome from a noiseless simulation of the implemented exchange sequence, with excellent agreement between the two.
	\figref{fig:heisenberg}{c} provides a schematic of how the entanglement wavefront created by a butterfly exchange with $\theta_B = \pi/2$ at the three selected butterfly locations spreads through the spin chain in time.
	Whenever the wave reaches one end of the chain, it reflects off the boundary but continues to travel at a fixed velocity.
	This is seen in both the experimental and noiseless results when $J_2 \approx 0$ ($\lambda_1 \approx 0$), with the evolution duration required for the first wave to ``break'' (i.e., cause a reduction in the singlet-return measurement) decreasing as the butterfly moves closer to the measurement site.
	Second, third, and fourth reductions in singlet-return probability are also observed.
	These features are caused by \emph{round-trip} reflections of the wavefront.

	The wavefront velocity reaches a minimum at $\lambda_1 = 0.5$.
	This is a special point known as the Majumdar-Ghosh (MG) model, where the analytic ground state of the infinite spin chain is made of spin-singlet dimers, just like our initial state.
	An OTOC is known to decay slowest when the initial state is near a ground state~\cite{Wei2019}, and we observe that behavior here even with our coarse Trotter step of $\Delta t J_1 = \pi/3$.
	Increasing $J_2$ complicates the Hamiltonian, modifying the velocity and dispersion of the wave.
	As $\lambda_1$ approaches unity, the highly connected Hamiltonian causes more complex dynamics, obscuring long-lived wave-like behavior.

	\section{Random OTOC-2}
	\label{sec:otoc-2}

	The second-order out-of-time-order correlator (\mbox{OTOC-2}) extends the OTOC-1 construction.
	As before, $M$ acts on a selected measured spin pair and $B$ on a selected butterfly spin pair.
	Whereas OTOC-1 applies the Heisenberg-picture butterfly operator $B(t)$ once, OTOC-2 applies it twice, separated by a physical $\pi$-exchange operation $M$ on the measured pair [smaller pink $M$ in \figref{fig:intro}{e}].
	The sequence applies $U(t)$, $B$, and $U^{-1}(t)$, then $M$, and repeats $U(t)$, $B$, and $U^{-1}(t)$ before singlet--triplet readout of the measured pair.
	
	In the implementation considered here, the intermediate $\pi$-exchange operation $M$ acts on the same spin pair that is measured at the end of the sequence.
	This choice follows the repeated-echo construction of Ref.~\onlinecite{Abanin2025} and allows the protocol to be written using a single measured-pair operator $M$.
	The intermediate exchange could instead act on a different pair, but we do not consider that generality here.
	Although here we report only the final singlet--triplet readout of the designated measured pair to be consistent with the repeated-echo construction, all spin pairs are read out at the end of each sequence and complete data is available.
	Using the singlet initialization and readout relation established for OTOC-1, the final singlet probability satisfies
	\begin{equation}
		\begin{aligned}
			1-p_S(t)
			&=
			\frac{1}{4}
			\bra{\psi_0}
			[B(t)M B(t),M]^\dagger
			\\
			&\qquad {}\times
			[B(t)M B(t),M]
			\ket{\psi_0}.
		\end{aligned}
		\label{eq:otoc2_commutator}
	\end{equation}
	The squared-commutator structure is the same as for OTOC-1, but the single butterfly operator is replaced by the repeated echo $B(t)M B(t)$.
	We note that if $B$ is also Hermitian and unitary, then $B^\dagger(t)=B(t)$, and the corresponding fourth-order correlator reduces to $\mathcal{C}_{\psi_0}^{(4)}(t)=\langle(B(t)M)^4\rangle_{\psi_0}$, the form used in Ref.~\onlinecite{Abanin2025}.
	\mbox{OTOC-2} is fourth-order in the evolved butterfly operator and sensitive to additional correlations and interference between its spreading components, making it more challenging for classical simulation~\cite{Abanin2025}.

\begin{figure}
		\centering
		\includegraphics[width=\columnwidth]{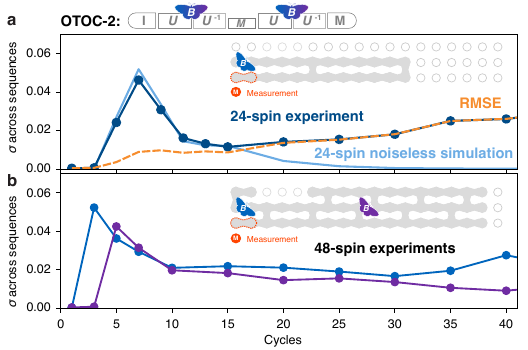}
		\caption{
			\label{fig:otoc2}
			\textbf{a} Standard deviation in singlet-return probability across different random sequences ($\sigma$) for 24-spin OTOC-2 as a function of cycle depth in the random unitary $U$. 
			Experiment is shown in dark blue and noiseless simulation in light blue.
			The root-mean-square error (RMSE) between the measured singlet-return probabilities and those predicted by the simulation can be evaluated in the same units as $\sigma$ and is shown in orange.
			The inset shows the 54-spin device with the active 24-spin region in gray, the butterfly location in blue, and the measurement site outlined in red.
			\textbf{b} Same as (a), but now for 48-spin random OTOC-2 at two different butterfly sites (blue and purple).
			In this case, simulation of the noiseless sequence is computationally intensive, so no RMSE is reported.
		}
	\end{figure}

	We first look at a 24-spin experiment and compare it to a noiseless simulation.
	\figref{fig:otoc2}{a} shows the 24-spin layout, including the location of the butterfly exchange $B$ and measurement site $M$.
	In this layout, the first vertical exchange electrode between the bottom two rails is omitted to allow for a faster pulsing cadence.
	By sweeping the number of cycles in $U$ across 100 random instances, we report in \figref{fig:otoc2}{a} the standard deviation $\sigma$ of the shot-averaged singlet-return probability.
	We use 10,000 shots per sequence, setting a Poisson shot-noise floor for $\sigma$ of $(10,000)^{-1/2} = 0.01$.
	
	To quantify the performance of the OTOC-2 experiment, we compute the expected $\sigma$ from a noiseless simulation of the same exchange sequences in the infinite-shot limit.
	We see good agreement in $\sigma$ up to 15 cycles, after which the $\sigma$ from the noiseless simulation begins to decay to zero while the experimentally measured $\sigma$ rises slowly to 3\%.
	To quantify the fidelity more precisely, we also report the root-mean-square error (RMSE) between the shot-averaged experimental result and the ideal outcomes:
	\begin{equation}
		\textrm{RMSE} = \sqrt{\frac{1}{N_U} \sum_{i=1}^{N_U} (p_\textrm{exp}(i) - p_\textrm{sim}(i))^2},
	\end{equation}
	where $p_\textrm{exp}(i)$ is the experimentally obtained return probability for the $i$th sequence, $p_\textrm{sim}(i)$ is the return probability of the noiseless simulation, and $N_U = 100$ is the number of random instances of $U$.
	Below 15 cycles, the RMSE is less than the measured $\sigma$, indicating that our device is providing an accurate, sequence-dependent signal.
	For this 24-spin layout, there are 24 active axes, and thus 24 exchanges in $U$ per cycle. 
	The total number of non-butterfly exchange pulses at 15 cycles is thus $15 \times 24 \times 4 = 1440$ exchange pulses, where the $4$ is due to the appearance of $U$ and $U^{-1}$ twice each in the OTOC-2 sequence.
	Setting $\epsilon_F = 3.2 \times 10^{-4}$, a naive linear estimate of the \mbox{OTOC-2} infidelity on our device would be $1440 \times \epsilon_F \approx 0.43$. 
	Thus, the OTOC-2 performance becomes unreliable at the sequence depth predicted by the $\epsilon_F$ extracted from XEB and MRB.
	
	Experimentally measured $\sigma$ from OTOC-2 on 48 spins with two different butterfly exchanges is shown in \figref{fig:otoc2}{b}.
	Despite sharing the same location for the butterfly exchange, the peak in $\sigma$ for the 48-spin blue butterfly occurs twice as early as that in the 24-spin case due to inclusion of a vertical exchange axis which halves the distance between the butterfly and measurement site.
	Now there are 54 pulses per cycle in $U$, and we expect the observed $\sigma$ to be representative of the noiseless sequence outcomes up to $(2 \times \epsilon_F \times 54 \times 4)^{-1} \approx 7$ cycles.
	
	\section{Discussion}
	\label{sec:discussion}

	Recent OTOC-2 experiments on a superconducting quantum computer presented a robust claim of verifiable quantum advantage~\cite{Abanin2025}.
	We do not make the same claim here: the dimension of the Hilbert space accessed is smaller, and efficient classical algorithms for $S=0$ exchange sequences are yet to be explored.
	The crux of the quantum-advantage argument for OTOC-2 in Ref.~\onlinecite{Abanin2025} was the emergence of second-order loop interference patterns in the final states' Pauli strings.
	In the 48-spin $S=0$ subspace, we do not have access to local Pauli operators, but we recently constructed an equivalent argument by replacing the Pauli strings with the more general Weyl-Heisenberg operators~\cite{Hoffman2026}.
	It is possible that the 48-spin OTOC-2 experiment (Fig.~\ref{fig:otoc2}) is the first demonstration of quantum advantage on an exchange-only architecture, but validating this claim requires thorough analysis of classical approaches versus experimental accuracy.

	A time-discretized $J_1$-$J_2$ Heisenberg spin chain was simulated on an 18-spin layout via an OTOC computation.
	Clear round-trip reflections of the entanglement wavefront generated by the butterfly are evident, and their durations depend on $J_2$, showcasing the high performance of the device.
	However, even with finer time discretization, this calculation is not yet large enough to provide quantitative predictions of the phases of an infinite spin chain.
	As constraints on exchange fidelity scale quadratically with the spin count, $\epsilon_F \propto N^{-2}$, to obtain similar quality results on a 180-spin chain, we would need $\epsilon_F \approx 3 \times 10^{-6}$!
	This highlights how a fully digital quantum computer requires quantum error correction to provide useful simulations of physical problems.
	A non-fault-tolerant device must avoid Trotter errors entirely, either by using analog quantum operations~\cite{Andersen2025,Andersen2026} or by targeting a time-dependent Hamiltonian~\cite{Leviatan2026}.

	The $\epsilon_F = 3.2 \times 10^{-4}$ measured on the 48-spin singlet manifold is an order of magnitude lower than state-of-the-art error rates on other platforms, which are all above $10^{-3}$~\cite{Chen2024, Madzik2025, Abanin2025, Cosentino2026, Dasu2026, AtomComputing2026, Evered2026}.
	With exchange-only control, computational qubits must be encoded in collective states of multiple spins, so qubit operations require compiled sequences of multiple exchange pulses.
	In the simplest picture of additive errors, this pulse-count overhead introduces a multiplicative factor when relating the error per exchange operation to the error per exchange-only qubit operation.
	For instance, applying $\ts\epsilon{F}$ to a 37-exchange-pulse CNOT yields an error of $37 \times 3\times10^{-4}\approx 1\%$, which is in the range of gate-level error attributions on a similar QPU \cite{tn3_paper}.  
	However, this additive error model is very rough, especially because the errors observed here include coherent contributions. 
	The roughness of this estimation highlights the value of these system-level benchmarks: error is more, and sometimes less, than the sum of its parts.  
	In particular, component-level measurements and microscopic noise modeling of our system predict an exchange infidelity approximately an order of magnitude below that observed in our system benchmarks, suggesting that further reductions of error are possible through improvements in control hardware, pulse delivery, calibration, and operation of the cryogenic controller.

	The work presented here provides a glimpse into alternative fault-tolerant architectures for EO quantum computing.
	Although three-spin DFS qubit encodings provide a universal gate set, there is no guarantee that they will be the best choice for every algorithm and application.
	One could search for significant performance gains by defining spin groupings of various sizes, each to play specialized computational roles in different parts of a future fault-tolerant circuit. 
	The high and improving fidelity of exchange-only operation leaves many options open for future computational tasks in this modality.
	
	\begin{acknowledgments}
	The authors acknowledge contributions from Liam C. O'Brien, Skylar Turner, and Jack T. Wilson.
	\end{acknowledgments}

	\appendix

	\section{State-preparation and measurement}
	\label{appendix:spam}
	
	The native state preparation and measurement (SPAM) of an exchange-only device consists of the creation and measurement of electron spin pairs in the singlet state.
	To achieve an initial $S=0$ global state on 48 spins, we must prepare 24 singlet pairs.  
	This preparation begins at the start of each XEB, MRB, or OTOC experiment as described in the manuscript, with a single electron in each dot in the device.
	This charge state of a single electron per dot is established when the device is first tuned and persists throughout many experiments, except when single electrons are ``flushed'' into baths and restored as singlets during singlet initialization at SPAM sites~\cite{tn3_paper}. 
	
\begin{figure*}
		\centering
		\includegraphics[width=\linewidth]{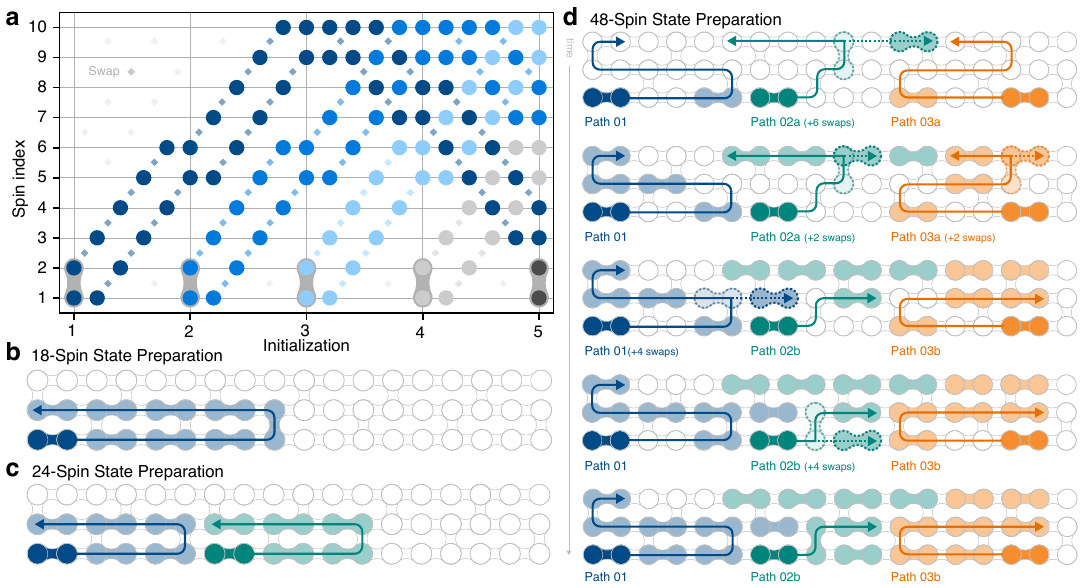}
		\caption{
			\label{fig:spam}
			\textbf{a} Example of a conveyor-belt $S=0$ SPAM technique for a 1D device with ten spins.
			The vertical axis indexes spin positions (1 through 10), while the horizontal axis indexes instructions.
			The instructions are of two types: initialization or spin swaps ($\pi$-exchanges), with five layers of swaps between successive initializations. 
			During spin-pair initialization, a pair of spins with a unique color is added to the figure, and that color is tracked through the figure to help visualize where each spin's initialization originates.
			Measurement of these spins uses the same spin-swap instructions, but the spin-pair initialization is replaced with measurement.
			\textbf{b} 18-spin SPAM, using a single SPAM site (dark blue) and a single conveyor belt path (blue arrow).
			\textbf{c} 24-spin SPAM, using two SPAM sites (dark blue, dark green) and two paths (blue and green arrows).
			\textbf{d} 48-spin SPAM, using three SPAM sites (dark blue, green, and orange) and three paths. 
			Groups of initializations occur in order from the top to the bottom of the figure.
			The groups are separated by either a path modification (solid arrow) or the presence of bespoke swap instructions (dashed arrows), which move a single spin pair to a location not on the current path.
		}
	\end{figure*}

	In the following, we describe a memory-efficient $S=0$ SPAM method, which we call a ``conveyor belt'' of spin pairs.  
	We emphasize that this differs significantly from ``conveyor-mode shuttling'' in spin-qubit devices~\cite{seidler_conveyor-mode_2022, de_smet_high-fidelity_2025}, since dots are never empty; what is ``conveyed'' is a well-defined spin state via swapping among always-occupied dots, not motion of charge between empty dots.
	In the spin-pair conveyor-belt, we 
	\begin{enumerate}
		\item Assign the \(N\) spins to neighboring spin pairs, which will each hold a singlet state after initialization. 
		\item Pick which SPAM sites on the device will be used for initialization and measurement (for now, assume the same sites are chosen for both). 
		\item Route a set of noncrossing ``paths'' that extend from each SPAM site, travel through spins along exchange axes, and eventually cover all spins \emph{without} crossing each other. 
		\item Run a	``conveyor-belt'' routine of \(\pi\)-pulses (two per exchange axis)	along each path after each initialization (or before each measurement), which moves a singlet pair down its path by four spins, as depicted in \figref{fig:spam}{a}.
	\end{enumerate}
	
	For step 3, ideally the paths have no branches, and each path is the same length. 
	In practice, with a finite yield of performant axes, this is not always possible, and so some device-specific customization of the otherwise automated method listed above is necessary. 
	This customized SPAM routing solution is achieved by introducing subroutines with varying path lengths to cover dead-end branches, along with small spin-swapping routines in some cases, as shown in \figref{fig:spam}{d}.
	In general, the customized SPAM routines are detrimental to SPAM performance (due to greater routing overhead and ``idling'' singlet pairs, which we partially address with dynamical decoupling) and use more cryo-controller memory.
	This technique is also visible in the ``Init'' and ``Measure'' sections of \figref{fig:intro}{c}.
	
	Note that step 4 does not move a singlet pair (made of two spins) over two spins but rather over four spins.
	This means that after a few applications of the initialization routine, the singlet pairs can be thought of as ``traveling'' along each path, but will be separated from one another by two uninitialized spins. 
	When a singlet pair reaches the end of the path, the structure of the \(\pi\)-pulses will cause its motion to reverse. 
	Viewing this motion as a wavefront propagation problem, the singlet pairs ``reflect'' off the end of the path and begin to travel back toward the SPAM site. 
	During this return trip, the singlet pairs now populate those uninitialized spins in between the singlet pairs traveling away from the SPAM site.
	
	\section{Generating random sequences}
	\label{appendix:rcs}
	
\begin{figure*}
		\centering
		\includegraphics[width=\linewidth]{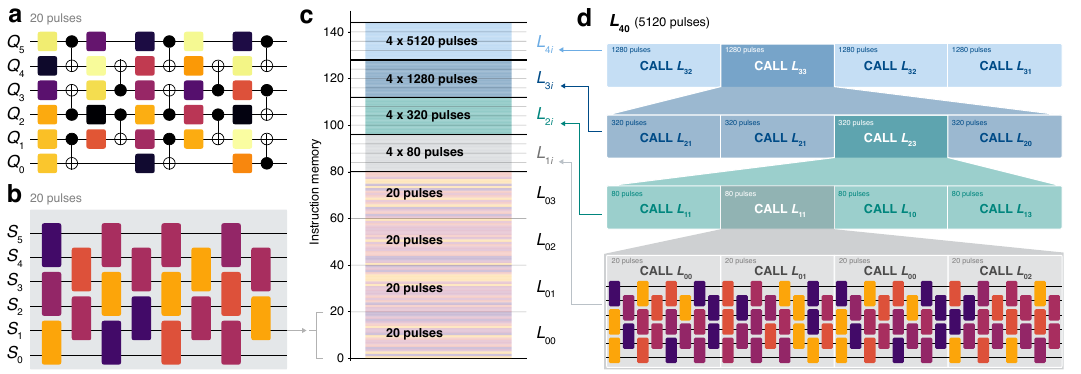}
		\caption{
			\label{fig:rcs}
			\textbf{a} Brickwork pattern for random circuit sampling on six qubits with CNOT gates as fixed entanglers. 
			The colored squares are random single-qubit rotations.
			\textbf{b} $S=0$ brickwork pattern for random sequence sampling on six spins.
			Each exchange has an angle randomly sampled from five values (color).
			This shows four cycles of exchanges (20 exchange pulses total).
			\textbf{c} Cryo-controller instruction memory usage during recursive random sequence compilation. 
			Here we set $b=4$, so there are four cycles per lowest-layer subroutine ($L_{0i}$ with $i \in [0,3]$), leading to twenty pulses in each base-layer subroutine, as in (b).
			Every other subroutine implements four calls to the layer below it, costing only four instructions but executing exponentially more pulses with only linear instruction-memory cost.
			\textbf{d} Example of the recursive subroutine compilation method for random brickwork $S=0$ sequences, showing an element of the fourth recursive layer, $L_{40}$, which executes 5,120 exchange pulses.
		}
	\end{figure*}

	To perform a random sequence on the $S=0$ subspace, an exchange is performed with a random angle \(\theta_i\).
	These exchanges can be organized into ``cycles,'' in which every available exchange is activated exactly once across the device [\figref{fig:rcs}{b}].
	As the cryo-controller has limited memory, after the instructions required for 48-spin SPAM are accounted for, only about 1,000 such pulse instructions can be played if each is uniquely defined in the program.
	If we assign one instruction to each pulse, the above sequence can only reach the asymptotic limit of zero fidelity if \(\epsilon_{F} > 10^{-3}\). 
	To accurately measure gate errors smaller than this, we will need a more memory-efficient sequence generation technique.
	
	The solution we have chosen is the use of recursive subroutine calls. 
	This allows for exponential growth in the number of cycles with instruction memory, instead of linear growth [\figref{fig:rcs}{c,d}].
	We first generate $b$ random sequences of $b$ cycles each, denoted $L_{0i}$ for $i=0,\ldots,b-1$, which form the lowest layer of the recursive construction.
	At each subsequent layer $\ell$, we construct $b$ subroutines $L_{\ell i}$, each comprising $b$ calls to randomly selected subroutines from layer $\ell-1$.
	Thus, each subroutine at layer $\ell$ contains $b^{\ell+1}$ cycles.
	We repeat this construction as needed and assemble the final program by randomly calling subroutines from the appropriate layers to reach the desired cycle depth.
	
	For example, if $b=4$, each $L_{0i}$ applies four cycles of random exchanges, each $L_{1i}$ applies sixteen cycles, etc. 
	To apply $68$ ($64+4$) cycles, we simply call one random $L_{2i}$ subroutine, and then one $L_{0i}$ subroutine, costing two instructions of memory. 
	If we assume each cycle has $5$ pulses, then the lowest layer costs $5 \times 4 \times 4 = 80$ instructions, and each subsequent layer costs $4 \times 4 = 16$ instructions.
	With only four layers, a single instruction can play over $5,000$ exchanges, as shown in~\figref{fig:rcs}{c}.

	In practice, we have chosen $b=7$ to maximize the subroutine randomness while still reaching the desired maximum depth ($10^{4}$ pulses) within controller memory.
	For $b=7$, the efficiency of this recursive method is similar to that of fully random selection for each pulse.
	The nested nature of this random sequence sampling technique is evident in the repeated colored patterns visible in the $U$ and $U^{-1}$ sections of \figref{fig:intro}{c}.
	
	\section{Component performance}
	\label{appendix:characterization}
	
\begin{figure*}
		\centering
		\includegraphics[width=\linewidth]{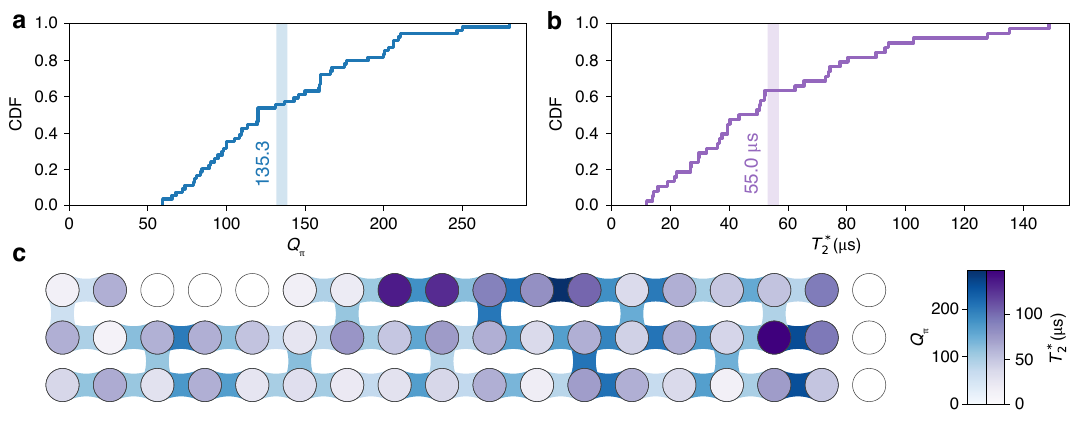}
		\caption{
			\label{fig:device_char}
			Microscopic characterization of the QPU used in this work.
			\textbf{a} Cumulative distribution function (CDF) for $Q_{\pi}$, which characterizes the error accumulated during exchange. The mean value is given by the vertical line.
			\textbf{b} Cumulative distribution function (CDF) for $T_2^*$, which characterizes the error accumulated during idle periods. The mean value is again given by the vertical line.
			\textbf{c} Colormap of $Q_{\pi}$ values in blue, plotted on exchange axes, and $T_2^*$ values in purple, plotted on spins.
		}
	\end{figure*}

	In Fig.~\ref{fig:device_char}, the axis-specific charge noise and dot-specific magnetic noise are summarized.
	$Q_{\pi}$ measures how many $\pi$-exchange pulses can be performed between spins $B$ and $C$ before a singlet state prepared on spins $A$ and $B$ decoheres.
	Measurement of a related quantity, $N_{osc}$, in a similar QPU but with room-temperature control electronics provided a mean value of 674~\cite{tn3_paper}.
	As the pulse sequence used to measure $Q_{\pi}$ is created by the cryo-controller, it can also be affected by pulse errors.
	For this reason, the $Q_{\pi}$ values (mean of 135) are likely bounded by pulse errors and not charge noise on the qubit chip itself.
	Alongside differences in the control electronics, the increase in measured exchange error could also be caused by additional opportunities for spin or charge state transitions during the rise and fall of each pulse:
	the measurement of $Q_{\pi}$ (a ``Pulsed Rabi'' experiment) uses thousands of individual pulses, while the measurement of $N_{osc}$ (a ``Rabi'' experiment) uses a single, long-duration pulse.
	
	$T_2^*$ measures the characteristic lifetime of a singlet pair while it sits idle, and is usually set by magnetic gradients between dots, both from static magnetic fields and dynamic polarization of magnetic nuclei in the SiGe quantum well.
	Here we report it in terms of spins, not axes, as we have used a matrix inversion to transform a set of axis-wise $T_2^*$ measurements into single-dot $T_2^*$ values via the relationship
	\begin{equation}
		\label{eq:t2star}
		\left( T_{2,ij}^* \right)^{-2} = \frac{1}{2} \left( \left( T_{2,i}^* \right)^{-2} + \left( T_{2,j}^* \right)^{-2} \right),
	\end{equation}
	where $T_{2,ij}^{*}$ is the measured decoherence time of an axis spanning dots $i$ and $j$, with $T_{2,i}^*$ and $T_{2,j}^*$ being spin-specific attributions of that decoherence mechanism.
	The mean $T_2^*$ value here is $2.5$ times higher than that reported previously for a similar QPU (19.3 $\mu$s)~\cite{tn3_paper}.
	This is due to improvements in the uniformity of static magnetic fields on the qubit chip and in the epitaxy of the quantum well.
	
	The singlet- and triplet-assignment infidelities~\cite{Blumoff2022} for the three SPAM sites on the QPU used in this work are presented in Table~\ref{tab:spam}.
	
	\begin{table}
		\begin{tabular}{ | r | l | l | }
			\hline
			SPAM site & Singlet & Triplet \\
			\hline
			Left & $6.0 \times 10^{-4}$ & $6.0 \times 10^{-3}$ \\
			Middle & $2.6 \times 10^{-3}$ & $1.6 \times 10^{-2}$ \\
			Right & $6.2 \times 10^{-3}$ & $7.8 \times 10^{-3}$ \\
			\hline
		\end{tabular}
		\caption{
			Assignment infidelities for singlet and triplet state preparation and measurement for the three SPAM sites used throughout this work (red dashed spin pairs in~\figref{fig:intro}{b}).
			The assignment infidelity is given by $1 - P[x,x]$, where $P[x,y]$ is the probability of measuring state $x$ after preparing state $y$, with $x=0$ for singlet and $x=1$ for triplet.
			}
		\label{tab:spam}
	\end{table}
	
	\section{Purity analysis}
	\label{appendix:xeb}

	\begin{figure}
		\centering
		\includegraphics[width=1\columnwidth]{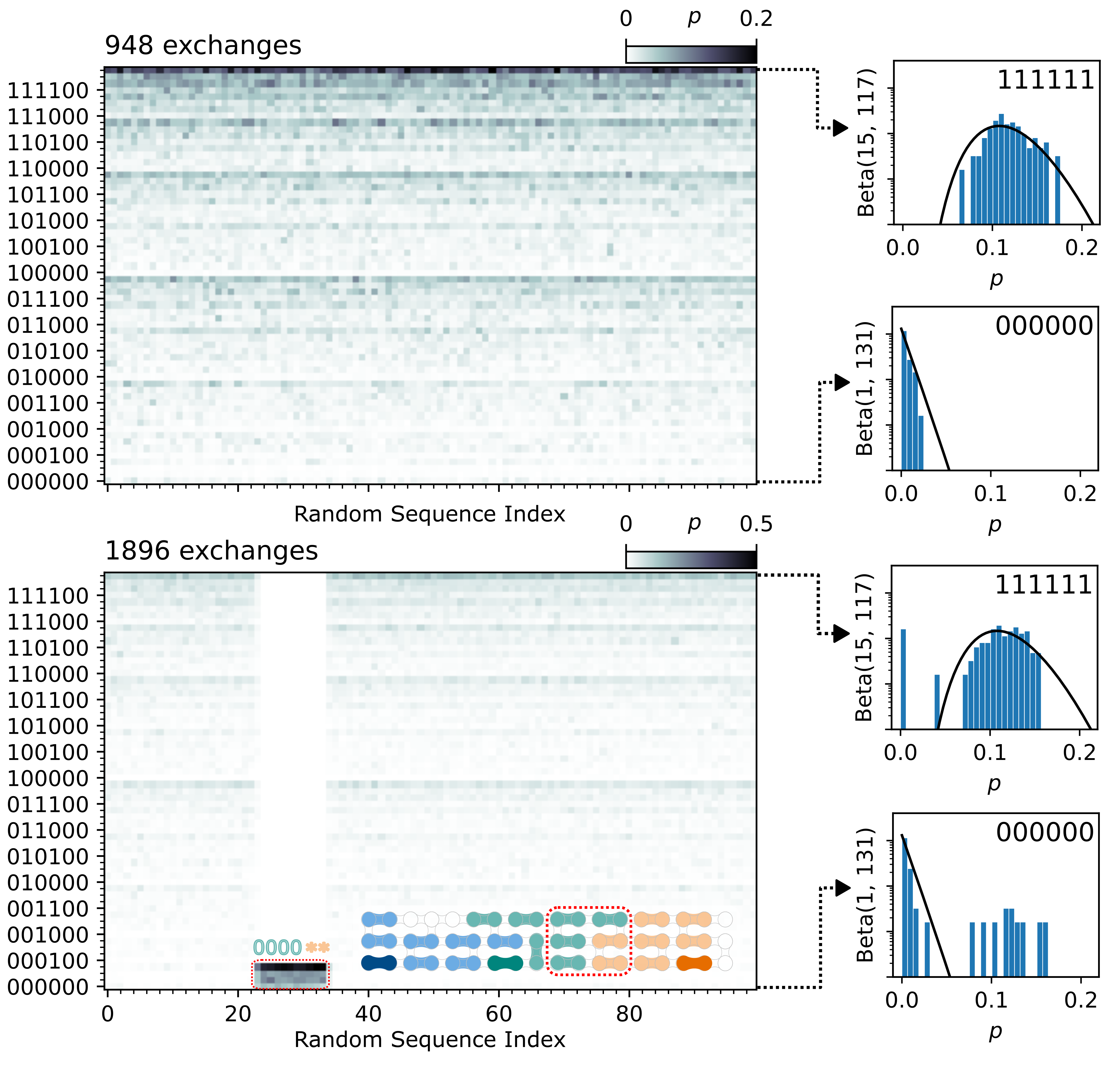}
		\caption{
			Probability ``speckles'' and measured output distributions for the sector of the $4\times 12$ XEB experiment in~\figref{fig:benchmarks}{b}.
			After 948 exchanges and across 100 random sequences, the output probabilities exhibit the expected speckle pattern across bitstrings and random sequence instances.
			Bitstrings containing exactly one triplet outcome (a single ``1'' in the bitstring) are forbidden by the total-spin-zero constraint and appear as white streaks across the sequence index, highlighting minimal leakage.
			For each bitstring, the histogram of measured probabilities over random sequence instances agrees with the corresponding $\mathrm{Beta}(R_{N_T},\Omega - R_{N_T})$ distribution (solid black curve) predicted for Haar-random states in the $S=0$ subspace with singlet--triplet measurement degeneracy.
			The all-zero bitstring is nondegenerate and follows $\mathrm{Beta}(1,131)$, the usual finite-dimensional Porter-Thomas distribution, whereas the all-one bitstring has maximal measurement degeneracy and follows $\mathrm{Beta}(15,117)$.
			At an exchange-pulse count of 1,896, some sequences experience correlated measurement error, corresponding to the rightmost gray data point in \figref{fig:benchmarks}{b}.
			Four of the six singlet pairs (green) are routed to one SPAM site which had a significant change in the shape and location of the PSB readout currents, causing all outcomes to register as singlets.
			The remaining two singlet pairs (orange) are routed to a second site and yield effectively random outcomes, visible as a population across the two least-significant bits. 
			This error occurs in 10 of the 100 random sequences, producing outliers outside the expected $\mathrm{Beta}$ distributions.
			These outliers destabilize the empirical variance relative to the Haar prediction and, consequently, the inferred purity.
			\label{fig:xeb_speckle}
		}
	\end{figure}
	
	We adapt speckle purity benchmarking (SPB)~\cite{Arute2019} to singlet--triplet readout using the Haar statistics outlined in Ref.~\cite{Hoffman2026}.
	In short, we bucket bitstring probabilities by the total number of triplets, $N_T$, and measure the variance in their probability histograms (right panels of Fig.~\ref{fig:xeb_speckle}).
	In the absence of incoherent noise, these histograms follow a Beta distribution.
	For simplicity, we do not attempt to directly compare the measured probability histograms to a Beta distribution and instead use the ratio of observed variance to the expected variance from the Beta distribution.

	To justify this approach, we now derive the connection between state purity and probability variance. 
	Let $p_{\mathbf{s}}(U;\rho)$ denote the probability of measuring bitstring $\mathbf{s}$ from the state $U\rho U^\dagger$.
	For pure $\rho$ and Haar-random $U$, the probability of measuring a bitstring containing $N_T$ triplets follows the distribution $\mathrm{Beta}(R_{N_T},\Omega-R_{N_T})$, where $\Omega$ denotes the dimension of the $S=0$ manifold. 
	The variance of the pure state's probabilities is therefore the variance of the Beta distribution: 
	\begin{equation}
		\frac{R_{N_T}(\Omega-R_{N_T})}{\Omega^2(\Omega+1)}.
		\label{eq:beta_variance}
	\end{equation}
	For mixed $\rho=\sum_{i=1}^{\Omega}\lambda_i|i\rangle\langle i|$, we have $p_{\mathbf{s}}(U;\rho)=\sum_i\lambda_i q_i(U)$, where $q_i(U)$ is the probability of the same bitstring for the rotated eigenstate $U|i\rangle$.
	All $q_i(U)$ have variance equal to that of the pure state (Eq.~\ref{eq:beta_variance}).
	By introducing $\Pi_\mathbf{s}$, which projects onto the states that correspond to the measurement bitstring $\mathbf{s}$, we obtain
	\begin{equation}
		\sum_i q_i(U) = \sum_i \bra{i} U^\dagger \Pi_\mathbf{s} U \ket{i} = \Tr(\Pi_\mathbf{s}) = R_{N_T}.
	\end{equation}
	Since $\sum_i q_i$ is constant, fluctuations in one element $q_i(U)$ must be compensated by all the others, and hence 
	\begin{equation}
		\mathrm{Cov}_U(q_i(U), q_j(U)) = - \frac{\mathrm{Var}_U(q_i(U))}{\Omega - 1}.
		\label{eq:covariance}
	\end{equation}
	Expanding the variance of the weighted sum therefore gives
	\begin{equation}
		\begin{aligned}
			&\mathrm{Var}_U[p_{\mathbf{s}}(U;\rho)]
			=
			\mathrm{Var}_U \left[ \sum_i \lambda_i q_i(U) \right]
			\\
			&= \sum_i \left[ \lambda_i^2 \mathrm{Var}_U(q_i(U)) + \sum_{j \neq i} \lambda_i \lambda_j \mathrm{Cov}_U(q_i(U), q_j(U)) \right]
			\\
			&= \mathrm{Var}_U[q_1(U)]
			\left[
				\sum_i\lambda_i^2
				-\frac{1-\sum_i\lambda_i^2}{\Omega-1}
			\right]
			\\
			&=
			\frac{R_{N_T}(\Omega-R_{N_T})}
				{\Omega^2(\Omega+1)}
			\frac{\Omega\mathrm{Tr}(\rho^2)-1}{\Omega-1},
		\end{aligned}
	\end{equation}
	where $\sum_i\lambda_i^2=\mathrm{Tr}(\rho^2)$.
	Note that, to obtain the third line, we used Eq.~\ref{eq:covariance} and the fact that, since $\sum_i \lambda_i = 1$, we have the identity $1 = (\sum_i \lambda_i)^2 = \sum_i \lambda_i^2 + \sum_{i \neq j} \lambda_i \lambda_j$.

	At each sequence length $n_{\mathrm{ex}}$, we compute the variance across random sequences for each allowed bitstring and then average these variances separately for the measured and ideal probabilities.
	Denoting the resulting quantities by $\mathrm{Var}_{\mathrm{obs}}$ and $\mathrm{Var}_{\mathrm{id}}$, we estimate the normalized purity as
	\begin{equation}
		\begin{aligned}
			\mathcal P(n_{\mathrm{ex}})
			&=
			\frac{\mathrm{Var}_{\mathrm{obs}}(n_{\mathrm{ex}})}
				{\mathrm{Var}_{\mathrm{id}}(n_{\mathrm{ex}})}
			\\
			&\approx
			\frac{\Omega\,\mathbb E_U\operatorname{Tr}
			[(\rho_U(n_{\mathrm{ex}}))^2]-1}{\Omega-1},
		\end{aligned}
	\end{equation}
	where $\rho_U(n_{\mathrm{ex}})$ is the mixed state for a fixed sequence $U$.
	Correlated readout errors can bias this estimate, as shown in Fig.~\ref{fig:xeb_speckle}.
	We fit $\mathcal P(n_{\mathrm{ex}})=A_{\mathcal P}(1-\epsilon_{\sqrt{\mathcal P}})^{2n_{\mathrm{ex}}}$, where $A_{\mathcal P}$ captures an initial offset arising from SPAM and routing.
	The factor of two allows for $\epsilon_{\sqrt{\mathcal P}}$ to match the XEB fidelity loss per exchange under depolarizing noise \cite{Arute2019}.

	\section{Trotter error of Heisenberg OTOC-1}
	\label{appendix:trotterization}

\begin{figure}
		\centering
		\includegraphics[width=1\columnwidth]{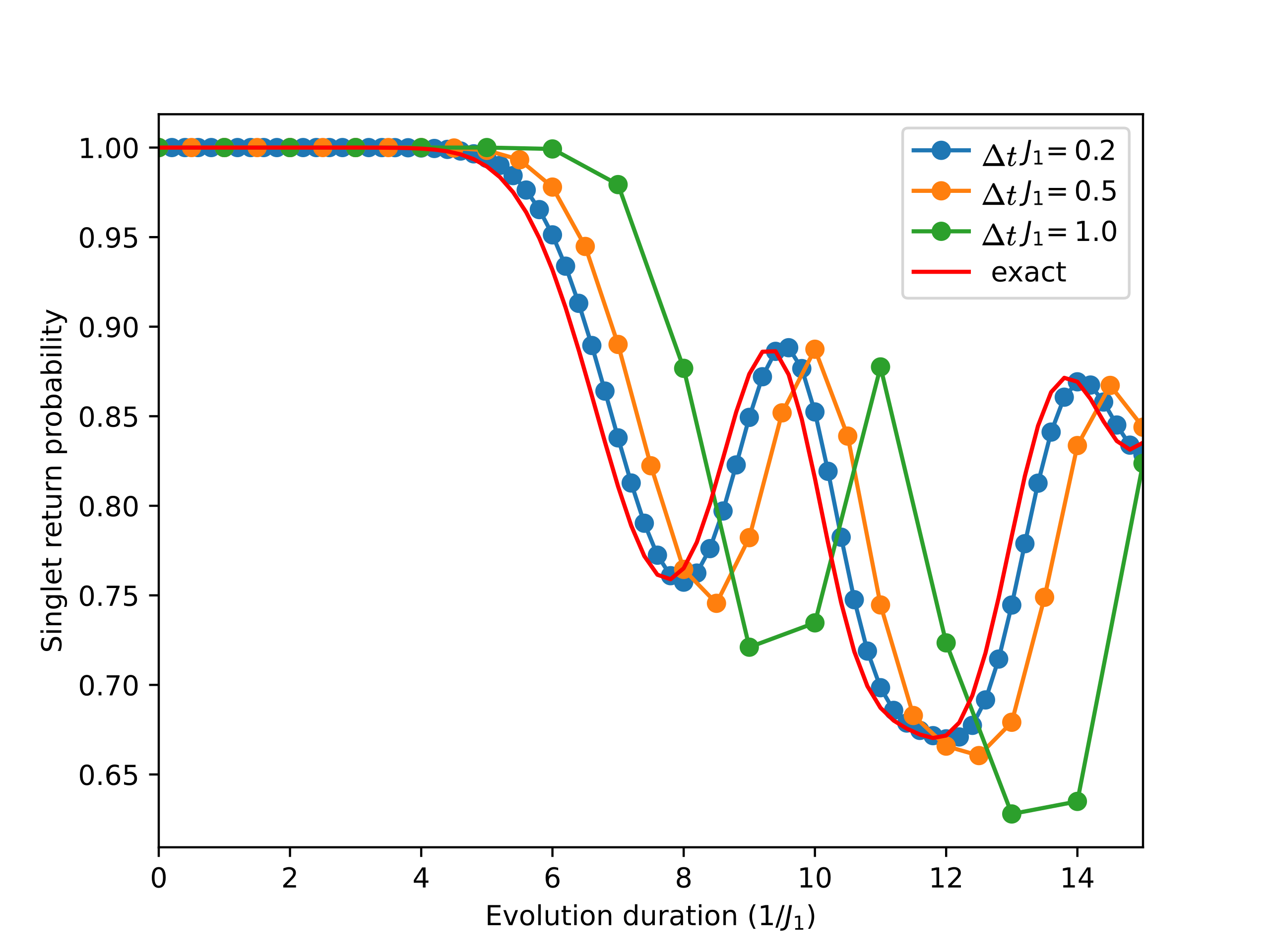}
		\caption{
			Exact calculation of the singlet-return probability from an OTOC-1 sequence on a 12-spin Heisenberg chain with $\lambda_1=0$ and $\theta_B=\pi/2$ (red). 
			Also shown are simulations of the NZ1-inspired schedule implemented in Sec.~\ref{sec:spin_chain} for three choices of Trotterization step size ($\Delta t J_1$).
     		\label{fig:trotter}
		}
	\end{figure}

	To estimate the Trotter error associated with the value of $\Delta t J_1 = \pi/3$, selected for the 18-spin Heisenberg experiment in Fig.~\ref{fig:heisenberg} of the main text, we perform simulations on a 12-spin Heisenberg chain (Fig.~\ref{fig:trotter}).
	In these simulations, the OTOC-1 is always evaluated at one end of the spin chain, with the $\pi/2$ butterfly exchange played on the opposite end.
	Although this chain is smaller than the experimental system and thus accrues less Trotter error, it illustrates that the results with $\Delta t J_1 = \pi/3 \approx 1$ are not a faithful reproduction of the physics of a $J_1$-$J_2$ spin-chain Hamiltonian.

	Because Trotter error scales linearly with the number of non-commuting interactions, if one wishes to keep the Trotter error constant for a larger spin system, one must reduce $\Delta t$ proportionally to the spin count $N$.
	And, as a large system requires more exchange operations per time step to apply the Hamiltonian, the number of operations to evaluate OTOC-1 for a single Trotter step grows like $N$.
	Combining these two effects, emulations of the Heisenberg chain with constant Trotter error require the number of exchanges to grow like $N^2$.
	So, compared to the results in Fig.~\ref{fig:trotter}, a 120-spin emulation requires a 100-fold decrease in the exchange error to simulate the same evolution duration with similar accuracy.
	For this reason, the use of a non-fault-tolerant digital quantum computer for Hamiltonian simulation is limited.

	\clearpage

	\hypertarget{hrl_authors}{$^\dagger$} 
	\textbf{HRL Quantum Team}\\
	Matt Abbitt$^{1}$,
	Michael Abraham$^{1}$,
	Edwin Acuna$^{1}$,
	I. Alverado$^{1}$,
	Carter Andrews$^{1}$,
	Hussein Anton$^{1}$,
	Katherine M. Beech$^{1}$,
	Aaron J. Bluestone$^{1}$,
	Jacob Z. Blumoff$^{1}$,
	Matthew G. Borselli$^{1}$
% ...

	\vspace{0.5em}
	\noindent
	$^1$ HRL Laboratories, LLC, Malibu, CA, USA \\
	$^2$ Microsoft Discovery \& Quantum, Redmond, WA, USA \\
	\mbox{\Envelope} email: stcarr@hrl.com

	\bibliography{refs}
	
\end{document}